# Method of characteristics and solution of DGLAP evolution equation in leading order at small-*x*

R BAISHYA[1,a], R RAJKHOWA[2,b] and J K SARMA[2,c]

[1]Physics Department, J N College, Boko-781123, Assam, India

[2]Physics Department, Tezpur University, Napaam-784028, Assam, India

E-mail: [a]rjitboko@yahoo.co.in; [b]rasna@tezu.ernet.in; [c]jks@tezu.ernet.in

**Abstract.** In this paper the singlet and non-singlet hadron structure functions have been obtained by solving Dokshitzer-Gribov-Lipatov-Alterelli-Parisi (DGLAP) evolution equations in leading order (LO) at the small-*x* limit. Here we have used a Taylor Series expansion and then the method of characteristics to solve the evolution equation. We have also calculated *t* and *x*-evolutions of deuteron structure function and the results are compared with the New Muon Collaboration (NMC) data.

## 1. Introduction

Deep Inelastic Scattering (DIS) [1-4] process is one of the basic processes for investigating the structure of hadrons. It is well known, all information about the structure of hadrons participating in DIS comes from the hadronic structure functions. According to QCD, at small value of *x* and at large values of $Q^2$ hadrons consist predominately of gluons and sea quarks, where *x* and $Q^2$ are Bjorken scaling variable and four momentum transfer in a DIS process respectively. The Dokshitzer-Gribov-Lipatov-Alterelli-Parisi (DGLAP) evolution equations give $t$ [= $ln\,(Q^2/\Lambda^2)$, $\Lambda$ is the QCD cut off parameter] and *x* evolutions of structure functions. Hence the solution of DGLAP evolution equations give quark and gluon structure functions which produce ultimately proton, neutron and deuteron structure functions.

Among various methods for solving DGLAP evolution equations [5-8], in recent years, an approximate method suitable at small-*x* has been pursued with considerable phenomenological success [9-12]. That method is very simple and mathematically transparent. In that approach, the DGLAP equations are expressed as partial differential equations (PDE) in *x* and *t* using the Taylor series expansion of some structure functions valid to be at small-$x$ and particular solutions of the equations have been obtained by different arbitrary linear combinations of $U$ and $V$ of the general solution $f\,(U, V) = 0$. But one of the limitations of these solutions is that, as the evolution equations are PDE, their ordinary solutions are not unique solution, but a range of solutions, of course the range is very narrow. On the other hand, this limitation can be overcome by using method of characteristics [13-15].

The method of characteristics is an important technique for solving initial value problems of first order PDE. It turns out that if we change co-ordinates from $(x, t)$ to suitable new co-ordinates $(S, \tau)$ then the PDE becomes an ordinary differential equation (ODE). Then we can solve ODE by standard method. In figure 1, the co-ordinates $(S, \tau)$ are considered such that the value of $S$ changes along a vertical curvy line where $\tau$ is constant and $\tau$ changes along a horizontal curvy line where $S$ is constant. For $t$-evolution, we consider that $S$ changes along the characteristics curve $[x(S), t(S); 0<S<\infty]$ and $\tau$ changes along the initial $(t = t_0)$ curve. On the other hand, for $x$-evolution, $\tau$ changes along the characteristics curve $[x(\tau), t(\tau); 0< \tau <\infty]$ and $S$ changes along the initial curve $(x = x_0)$. Along the characteristics curve we get one ODE with one boundary condition. After solving it and transforming $(S, \tau)$ again to $(x, t)$ we get the unique solution.

In this paper, we obtain a solution of DGLAP equations for singlet and non-singlet structure functions at small-$x$ at LO by using this method of characteristics. The result is compared with NMC data [16] for deuteron structure function. Here the section 1 is the introduction, section 2 deals with the necessary theory and section 3 is the results and discussion.

## 2. Theory

The DGLAP evolution equations for singlet and non-singlet structure functions in LO in standard form are [17]

$$\frac{\partial F_2^S}{\partial t} - \frac{A_f}{t}\left[\{3 + 4\ln(1-x)\}F_2^S(x,t) + I_1^S(x,t) + I_2^S(x,t)\right] = 0\ , \tag{1}$$

$$\frac{\partial F_2^S}{\partial t} - \frac{A_f}{t}\left[\{3 + 4\ln(1-x)\}F_2^{NS}(x,t) + I_1^{NS}(x,t)\right] = 0\ , \tag{2}$$

where

$$I_1^S(x,t) = 2\int_x^1 \frac{dw}{1-w}\left[\left(1+w^2\right)F_2^S\left(\frac{x}{w},t\right) - 2F_2^S(x,t)\right], \tag{3a}$$

$$I_2^S(x,t) = 2N_f\int_x^1 \left\{w^2 + \left(1-w^2\right)\right\}G\left(\frac{x}{w},t\right)dw, \tag{3b}$$

$$I_1^{NS}(x,t) = 2\int_x^1 \frac{dw}{1-w}\left[\left(1+w^2\right)F_2^{NS}\left(\frac{x}{w},t\right) - 2F_2^{NS}(x,t)\right], \tag{3c}$$

where $A_f = \frac{4}{33-2N_f}$ and $N_f$ being the flavors number.

Let us introduce the variable $u = 1\text{-}w$ and note that

$$\frac{x}{w} = \frac{x}{1-w} = x\sum_{k=0}^{\infty} u^k . \tag{4}$$

Since $x<w<1$, so $0<u<1\text{-}x$ and hence the series (4) is convergent for $|u|<1$. Now using Taylor's expansionseries we can rewrite $F_2^S\ (x/w,\ t)$ and $G(x/w,\ t)$ as

$$F_2^S\left(\frac{x}{w},t\right) = F_2^S\left(x + x\sum_{k=1}^{\infty}u^k,t\right)$$

$$= F_2^S(x,t) + x\sum_{k=1}^{\infty}u^k\frac{\partial F_2^S(x,t)}{\partial x} + \tfrac{1}{2}x^2\left(\sum_{k=1}^{\infty}u^k\right)^2\frac{\partial^2 F_2^S(x,t)}{\partial x^2} + \ldots$$

$$\approx F_2^S(x,t) + x\sum_{k=1}^{\infty}u^k\frac{\partial F_2^S}{\partial x}, \tag{5a}$$

$$G\left(\frac{x}{w},t\right) \approx G(x,t) + x\sum_{k=1}^{\infty}u^k\frac{\partial G(x,t)}{\partial x}. \tag{5b}$$

Since $x$ is small in our region of discussion, the terms containing $x^2$ and higher powers of $x$ have been neglected. Using equation (5a) and (5b) in equation (3a) and (3b) and performing $u$-integrations we get

$$I_1^S(x,t) = -\{(1-x)(3+x)\}F_2^S(x,t) + \{2x\ln\left(\frac{1}{x}\right) + x(1-x^2)\}\frac{\partial F_2^S(x,t)}{\partial x}, \tag{6a}$$

$$I_2^S(x,t) = N_f\left[\tfrac{1}{2}(1-x)(2-x+2x^2)G(x,t) + \{-\tfrac{1}{2}x(1-x)(5-4x+2x^2)\}\frac{\partial G(x,t)}{\partial x}\right]. \tag{6b}$$

Putting equations (6a) and (6b) in equation (1) we get

$$\frac{\partial F_2^S(x,t)}{\partial t} - \frac{A_f}{t}\left[A(x)F_2^S(x,t) + B(x)\frac{\partial F_2^S}{\partial x} + C(x)G(x,t) + D(x)\frac{\partial G(x,t)}{\partial x}\right] = 0 \tag{7}$$

where

$A(x) = 3 + 4ln(1\text{-}x) - (1\text{-}x)(3 + x)\ @\ 2x,$ (8a)

$B(x) = x(1-x^2) + 2x\ln(1/x)$ @ $x$ , (8b)

$C(x) = (1/2)\ N_f\, x\ [(-1/2)(1-x)(5-4x + 2x^2) + (3/2)\ln(2x^3\,)]$ @ $(1/2)N_f(2-3x)$, (8c)

$D(x) = N_f x[(-1/2)(1-x)(5 - 4x + 2x^2\,) + (3/2)\ln(1/x)]$ @ $(-5/2)N_f\, x$ . (8d)

Here terms containing $x^2$ and higher powers of $x$ have been neglected as our first approximation. Now let us assume

$G\,(x, t) = k(x)\, F_2^S\,(x, t)$, (9)

where $k(x)$ is a suitable function of $x$ or may be a constant. We may assume $k(x) = k,\ ae^{bx}$, $cx^d$, where $k, a, b, c, d$ are suitable parameters which can be determined by phenomenological analysis.

Now equation (7) gives

$$\frac{\partial F_2^S\,(x,t)}{\partial t} = \frac{A_f}{t}\left[ L\,(x)F_2^S\,(x,t) + M\ (x)\frac{\partial F_2^S\,(x,t)}{\partial x}\right]$$

$$\Rightarrow\ -t\,\frac{\partial F_2^S\,(x,t)}{\partial t} + A_f\, L(x)F\ _2^S\,(x,\,t) + A_f\, M(x)\ \frac{\partial F_2^S\,(x,t)}{\partial x} = 0, \tag{10}$$

where

$L(x) = A(x) + k(x)\ C(x) + D(x)\ \partial\, k(x)/\ \partial\, x$, (11a)

$M(x) = B(x) + k(x)\ D(x)$. (11b)

For simplicity let us consider $k(x) = k$ (arbitrary constant). Then

$L(x) = A(x) + k\ C(x)$. (11c)

2(a). $t$-Evolution

Let us introduce two new variable $S$ and t instead of $x$ and $t$, such that

$$\frac{dt}{dS} = -t\,, \tag{12a}$$

$$\frac{dx}{dS} = A_f M(x), \tag{12b}$$

which are known as characteristics equations. As we can write

$$\frac{dF_2^S\,(x,t)}{dS} = \frac{\partial F_2^S\,(x,t)}{\partial t}\frac{dt}{dS} + \frac{\partial F_2^S\,(x,t)}{\partial x}\frac{dx}{dS} = \frac{dF_2^S\,(S,\tau)}{dS},$$

putting equations (12a) and (12b) we get

$$\frac{dF_2^S\,(S,\tau)}{dS} = -t\,\frac{\partial F_2^S\,(x,t)}{\partial t} + A_f M\,(x)\frac{\partial F_2^S\,(x,t)}{\partial x}.$$

Now from equation (10), we get

$$\frac{dF_2^S(S,\tau)}{dS} + L(S,\tau)F_2^S(S,\tau) = 0, \tag{13}$$

where

$$L(S,t) = A_f L(x) = A_f x(2 - (3/2)N_f k) + A_f N_f k. \tag{14}$$

Then equation (13) gives

$$\frac{dF_2^S}{dS} = -A_f [x(2 - \frac{3}{2}N_f k) + N_f k]F_2^S(S,\tau). \tag{15}$$

Now integrating equations (12a) and (12b) we get

$$ln(t) = -S + C_1, \tag{16a}$$

$$ln(x) = A_f(1 - \tfrac{5}{2}N_f k)S + C_2. \tag{16b}$$

Here $C_1$ and $C_2$ are constants of integration which can be obtained by applying boundary conditions. Let us consider the boundary condition: when $S = 0$, $t = t_0$, $x = t$. Hence equation (16a) gives

$$C_1 = ln(t_0) \text{ and } S = ln(t_0/t). \tag{17a}$$

Similarly equation (16b) gives

$$ln\left(\frac{x}{\tau}\right) = A_f\left(1 - \frac{5}{2}N_f k\right)S$$

$$\Rightarrow x = \tau\, exp[A_f\{1 - (5/2)N_f\ k\}S] \tag{17b}$$

$$\Rightarrow \tau = x(t/t_0)^{A_f\{1-(5/2)N_f k}\} \tag{17c}$$

$$\Rightarrow t = x(t/t_0)A_f\{1-(5/2)N_f k\}.$$

Using equation (17b) in equation (15) we get

$$\frac{dF_2^S(S,\tau)}{F_2^S(S,\tau)} = -A_f [(2 - (3/2)N_f k)\tau\, exp\{A_f(1 - (5/2)N_f k)S\} + N_f k]\ dS. \tag{18a}$$

Integrating equation (18a) we get

$$ln\,F_2^S(S,\tau) = -\frac{(2 - (3/2)N_f k)}{(1 - (5/2)N_f k)}\tau\, exp\{A_f(1 - (5/2)N_f k)S\} - A_f N_f k\, S + C_3, \tag{18b}$$

where $C_3$ is the constant of integration. Let us consider, when $S = 0$, $F_2^S(S, t) = F_2^S(t)$.

Then equation (18b) gives

$$C_3 = ln\,F_2^s(\tau) + \frac{(2 - (3/2)N_f k)}{(1 - (5/2)N_f k)}\tau \tag{19}$$

and

$$F_2^S(S,\tau)=F_2^S(\tau)\,exp\left[\frac{\left(2 - (3/2)N_f k\right)}{\left(1-(5/2)N_f k\right)}\tau\left[1 - exp\left\{A_f\left(1-(5/2)N_f k\right)S\right\}\right]-A_f N_f k\, S\right]. \qquad (20)$$

Now we are to replace the co-ordinate system *(S,* τ*)* by *(x, t)* applying equations (17a) and (17c) in equation (20),

$$F_2^S(S,\tau)=F_2^S(\tau)\,exp\left[\frac{\left(2-(3/2)N_f k\right)}{\left(1-(5/2)N_f k\right)}x\left(\frac{t}{t_0}\right)^{A_f\left(1-(5/2)N_f k\right)}\left[1 - exp\left\{A_f\left(1-(5/2)N_f k\right)ln\left(\frac{t_0}{t}\right)\right\}\right]-A_f N_f k\, ln\left(\frac{t_0}{t}\right)\right]$$

$$= F_2^S(\tau)\left(\frac{t}{t_0}\right)^{A_f N_f k} exp\left[\frac{\left(2-(3/2)N_f k\right)}{\left(1-(5/2)N_f k\right)}x\left(\frac{t}{t_0}\right)^{A_f\left(1-(5/2)N_f k\right)}\left\{1 - \left(\frac{t_0}{t}\right)^{A_f\left(1-(5/2)N_f k\right)}\right\}\right] \qquad (21)$$

When $S = 0$, $t = t_0$ and $x = \tau$; then the input function is $F_2^S(\tau)=F_2^S(x,t_0)$. Hence equation (21) gives

$$F_2^S(x,t)=F_2^S(x,t_0)\left(\frac{t}{t_0}\right)^{A_f N_f k} exp\left[\frac{\left(2-(3/2)N_f k\right)}{\left(1-(5/2)N_f k\right)}x\left\{\left(\frac{t}{t_0}\right)^{A_f\left(1-(5/2)N_f k\right)} - 1\right\}\right] \qquad (22)$$

which is the *t*-evolution of singlet structure function.

2(b). x-Evolution

Let us consider now, (*S,* τ) are variables such that

$$\frac{dt}{d\tau}=-t\,, \qquad (23a)$$

$$\frac{dx}{d\tau}=A_f M(x). \qquad (23b)$$

Hence, as before equation (10) gives

$$\frac{dF_2^S(S,\tau)}{d\tau}+L(S,\tau)F_2^S(S,\tau)=0\,, \qquad (24)$$

where *L(S,*τ*)* is defined by equation (14) and equation (24) gives

$$\frac{dF_2^S}{d\tau}=-A_f\,[x(2-\frac{3}{2}N_f k)+N_f k]F_2^S(S,\tau)\,.$$

Now equations (23a) and (23b) can be solved as

$$ln(t) = -\tau + C_4 \qquad (25a)$$

and

$$ln(x) = A_f(1-(5/2)\,N_f k)\,\tau + C_5. \qquad (25b)$$

Let us consider the boundary condition as, when $\tau = 0$, $x = x_0$ and $t = S$. So equation (25a) gives $C_4 = ln(S)$ and $\tau = ln(S/t)$ (26a)

Similarly equation (25b) gives $C_5 = ln(x_0)$ and

$$ln\left(\frac{x}{x_0}\right) = A_f\left(1-(5/2)N_f k\right)\tau$$

$$\Rightarrow x = x_0\, exp\left\lfloor A_f\left(1-(5/2)N_f k\right)\tau\right\rfloor, \tag{26b}$$

$$\Rightarrow \tau = \frac{1}{A_f\left(1-(5/2)N_f k\right)} ln\left(\frac{x}{x_0}\right). \tag{26c}$$

Equation (24) gives now

$$\frac{dF_2^S(S,\tau)}{F_2^S(S,\tau)} = -A_f\left[\left(2-(3/2)N_f k\right)x_0 exp\left\{A_f\left(1-(5/2)N_f k\right)\tau\right\} + N_f k\right]d\tau.$$

Integrating we get

$$ln\, F_2^S(S,\tau) = -\frac{\left(2-(3/2)N_f k\right)}{\left(1-(5/2)N_f k\right)} x_0\, exp\left\{A_f\left(1-(5/2)N_f k\right)\tau\right\} - A_f N_f k\,\tau + C_6. \tag{27}$$

Let us consider, when $\tau = 0$, $F_2^S(S,\tau) = F_2^S(S)$. So equation (27) gives

$$C_6 = \frac{\left(2-(3/2)N_f k\right)}{\left(1-(5/2)N_f k\right)} x_0 + lnF_2^S(S)$$

and

$$F_2^S(S,\tau) = F_2^S(S)\, exp\left[\frac{\left(2-(3/2)N_f k\right)}{\left(1-(5/2)N_f k\right)} x_0\left[1 - exp\left\{A_f\left(1-(5/2)N_f k\right)\tau\right\}\right] - A_f N_f k\,\tau\right]. \tag{28}$$

Now we can replace the co-ordinate system $(S, \tau)$ to $(x, t)$ by applying equation (26c) in equation (28)

$$F_2^S(S,\tau) = F_2^S(S)\, exp\left[\frac{\left(2-(3/2)N_f k\right)}{\left(1-(5/2)N_f k\right)} x_0\left\{1 - \frac{x}{x_0}\right\} + ln\left(\frac{x_0}{x}\right)^{(N_f k)/(1-(5/2)N_f k)}\right]$$

$$= F_2^S(S)\left(\frac{x_0}{x}\right)^{(N_f k)/(1-(5/2)N_f k)} exp\left[\frac{\left(2-(3/2)N_f k\right)}{\left(1-(5/2)N_f k\right)}\left(x_0 - x\right)\right]. \tag{29}$$

Here input function can be defined as, when $\tau = 0$, $x = x_0$, $t = S$ and

$$F_2^S(S) = F_2^S(x_0, t).$$

Hence equation (29) gives

$$F_2^S(x,t) = F_2^S(x_0,t)\left(\frac{x_0}{x}\right)^{(N_f k)/(1-(5/2)N_f k)} exp\left[\frac{(2-(3/2)N_f k)}{(1-(5/2)N_f k)}(x_0 - x)\right], \quad (30)$$

which is the $x$-evolution of singlet structure function. Proceeding in the same way, we get $t$ and $x$-evolutions of non-singlet structure function from equation (2) as

$$F_2^{NS}(x,t) = F_2^{NS}(x,t_0)\, exp\left[2x\left\{\left(\frac{t}{t_0}\right)^{A_f} - 1\right\}\right], \quad (31a)$$

$$F_2^{NS}(x,t) = F_2^{NS}(x_0,t)\, exp[2(x_0 - x)] \quad (31b)$$

respectively.

The deuteron, proton and neutron structure functions measured in DIS can be written in terms of singlet and non-singlet quark distribution functions as

$$F_2^d(x,t) = \frac{5}{2}F_2^S(x,t), \quad (32a)$$

$$F_2^p(x,t) = \frac{5}{18}F_2^S(x,t) + \frac{3}{18}F_2^{NS}(x,t) \quad (32b)$$

and

$$F_2^n(x,t) = \frac{5}{18}F_2^S(x,t) - \frac{3}{18}F_2^{NS}(x,t). \quad (32c)$$

The $t$ and $x$-evolution of deuteron structure functions can be obtained by putting equations (22) and (30) respectively in the equation (32a) as

$$F_2^S(x,t) = F_2^S(x,t_0)\left(\frac{t}{t_0}\right)^{A_f N_f k} exp\left[\frac{(2-(3/2)N_f k)}{(1-(5/2)N_f k)}x\left\{\left(\frac{t}{t_0}\right)^{A_f(1-(5/2)N_f k)} - 1\right\}\right], \quad (33a)$$

$$F_2^d(x,t_0) = F_2^d(x_0,t)\left(\frac{t}{t_0}\right)^{(N_f k)/(1-(5/2)N_f k)} exp\left[\frac{(2-(3/2)N_f k)}{(1-(5/2)N_f k)}(x_0 - x)\right], \quad (33b)$$

where

$$F_2^d(x,t_0) = \frac{5}{2}F_2^S(x,t_0), \quad (34a)$$

$$F_2^d(x_0,t) = \frac{5}{2}F_2^s(x_0,t). \quad (34b)$$

## 3. Result and discussion

In this paper, we compare our result of $t$ and $x$-evolutions of deuteron structure function $F_2^d$ measured by the NMC in muon-deuteron DIS with incident momentum *90, 120, 200, 280* $GeV$ [16]. Since the equation (22) and (31a) as well as (30) and (31b) are not in the same form, so we need to separate the input functions from the data points to extract the $t$ and $x$-evolution of proton and neutron structure function. So using equations (32b) and (32c), determination of evolutions of proton and neutron structure functions is not possible. For quantitative analysis, we consider the QCD cut-off parameter $\Lambda_{\overline{MS}} = 0.323\ GeV$ [18] for $\alpha_S\left(M_Z^2\right) = 0.119 \pm 0.002$ and $N_f = 4$. It is observed that our result is very sensitive to arbitrary constant $k$ in $t$-evolution and best fitting is in the range of $1.03 \le k \le 1.6$.

In figure 2 for $t$-evolution, we have plotted computed values of $F_2^d$ against $Q^2$ values for a fixed-$x$. In figure 2(a), we have plotted the graph for $x = 0.0045$ with $Q_0^2 = 0.75\ GeV^2$ as the initial point. The agreement of our result is found to be excellent at $k = 1.6$. Similarly in figures 2(b), 2(c) and 2(d) for $x = 0.008$, $x = 0.0125$ and $x = 0.0175$ respectively, the computed values are plotted against the corresponding values of $Q^2$ for the range $0.75\ GeV^2$ to $3.5\ GeV^2$. Here the input parameters are taken as for $Q_0^2 = 0.75\ GeV^2$ in first two curves and $Q_0^2 = 1.25\ GeV^2$ for the third curve. It is found that agreement of these results with data is excellent for the range $1.03 \le k \le 1.2$. In figure 2, the solid lines represent the best fitting curves. Except the best fitting curves, the dotted lines represent those for $k = 1.6$ and dashed lines represent for $k = 1.03$.

In figure 3 for $x$-evolution, we have plotted computing values of $F_2^d$ against the $x$-values for a fixed-$Q^2$. Here we have plotted the graphs for $Q^2 = 11.5, 15, 20$ and $27\ GeV^2$ for the range of $0.0025 \le x \le 0.14$. Here we have considered the input parameter $x_0 = 0.09$ for first three curves and $x_0 = 0.14$ for the fourth one. The best value of $k$ *is* $k = 0.01$. But as $Q^2$ increases the $k$-value slightly increases. For $Q^2 = 27\ GeV^2$, the excellent agreement will be found for $k = 0.02$. Here also the solid lines represent the best fit curves. Except best fit curves, the dotted lines represent the graphs for $k = 0.02$ and dashed line represents for $k = 0.01$.

Though there are various methods to solve DGLAP evolution equation to calculate quark and gluon structure functions, our method of characteristics to solve these equations is also a viable alternative. Though mathematically vigorous, it changes the integro-differential equations into ODEs and then makes it possible to obtain unique solutions. As our subsequent work we can calculate the $t$ and $x$-evolutions of nucleon structure functions by considering $k(x) = ax^b$ and $k(x) = ce^{dx}$. Also we will try to extend our work to next-to-leading order (NLO) at small-$x$.

## Acknowledgments

One of us (JKS) is grateful to UGC, New Delhi for the financial assistance to this work in the form of a major research project. We would like to acknowledge A K Bhattacharyya, A M P Hussain and B U Jamil of Tezpur University for valuable conversations.

## Appendix A

// Program for *t*-evolution of singlet structure function://

```
#include <math.h>
#include <stdio.h>
#include <conio.h>
#define Nf  4
#define k   1
#define z   0.323
#define Af 0.16
#define t  (log(Q2)-log(z*z))
#define t0  (log(Q02)-log(z*z))
#define A  Af*(1-2.5*Nf*k)
#define B  pow((t/t0),A)
#define D  (2-1.5*Nf*k)/(1-2.5*Nf*k)
#define E  pow((t/t0),(Nf*Af*k))
#define G  exp(D*x*(B-1))
#define F2(x,t) F0*E*G

main()
    {
    float Q2,Q02,x,F0;
    clrscr();
    printf("\nValue of Q2:");
    scanf("%f",&Q2);
    printf("\nValue of Q02:");
    scanf("%f",&Q02);
    printf("\nValue of x:");
    scanf("%f",&x);
    printf("\nValue of F0:");
    scanf("%f",&F0);
    printf("\nF2(x,t)=%f", F2(x,t));
    getch();
    return(0);
    }
```

## Appendix B

// Program for *x*-evolution of singlet structure function://

```
#include <math.h>
#include <stdio.h>
#include <conio.h>
#define Nf  4
#define k   1
#define z   0.323
#define Af 0.16
#define A  (Nf*k)/(1-2.5*Nf*k)
#define B  x/x0
#define C  (2-1.5*Nf*k)/(1-2.5*Nf*k)
#define D  C*(x0-x)
#define E  pow((x0/x),A)
#define G  exp(D)
#define F2(x,t) F0*E*G

main()
{
float x0,x,F0;
clrscr();
printf("\nValue of x0:");
scanf("%f",&x0);
printf("\nValue of x:");
scanf("%f",&x);
printf("\nValue of F0:");
scanf("%f",&F0);
printf("\nF2(x,t)=%f", F2(x,t));
getch();
return(0);
}
```

## Figure captions

**Figure 1.** Characteristic curves for constant values of $t$ ($t_1$, $t_2$, $t_3$ . . .). The values of $S$ change along a vertical characteristic curve. On the other hand, and along a horizontal characteristic curve, the values of $t$ change for constant values of $S$ ($S_1$, $S_2$, $S_3$ . . .) in the $t$-$x$ plane.

**Figure 2.** Results of $t$-evolution of deuteron structure function $F_2^d$ for the given value of $x$. Data points at lowest-$Q^2$ values in the figures are taken as input to test the evolution equation (22).

**Figure 3.** Results of $x$-evolution of deuteron structure function $F_2^d$ for the given values of $t$. Data point at lowest-$x$ values in the figures are taken as inputs to test the evolution equation (30).

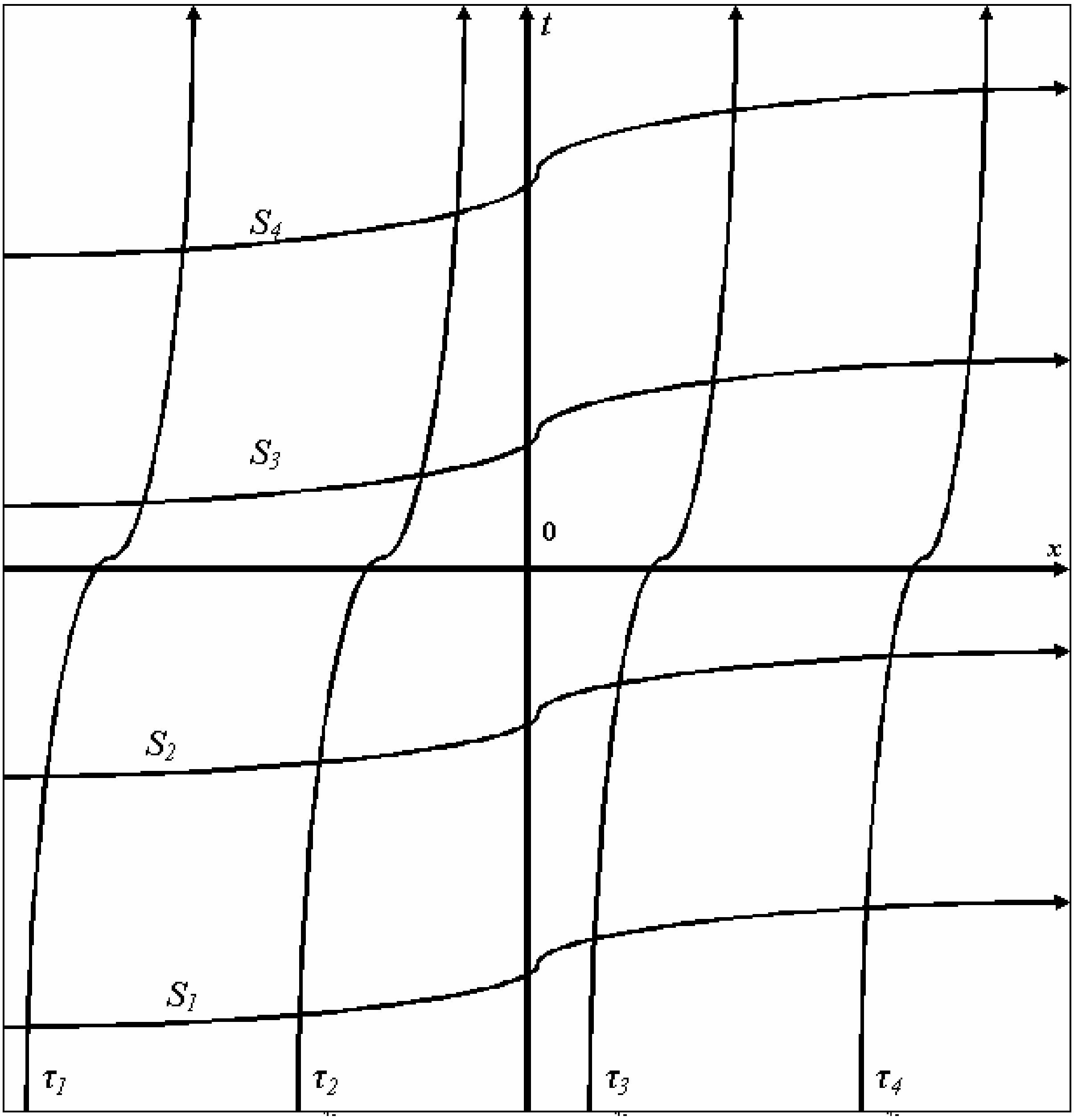

Figure 1

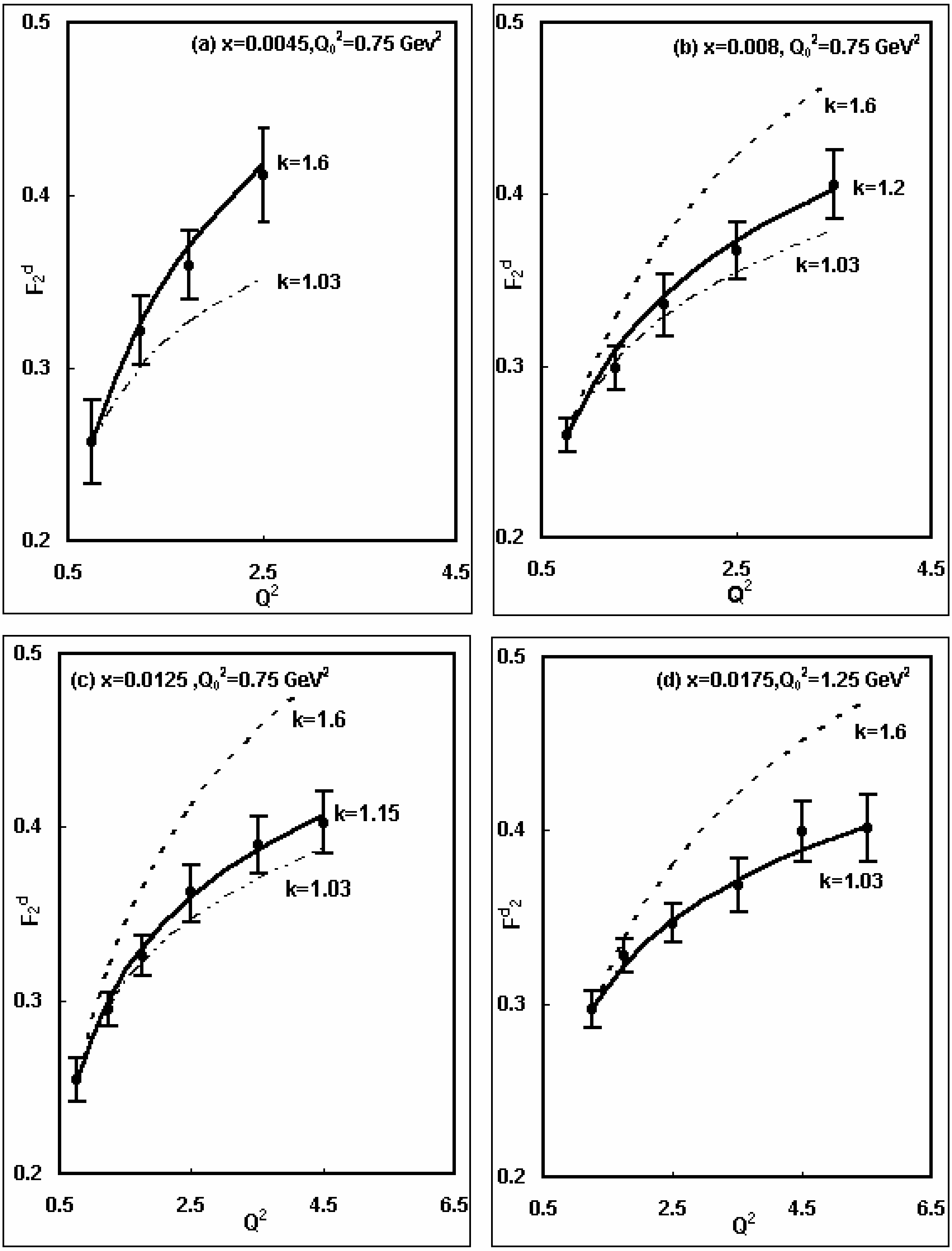

Figure 2

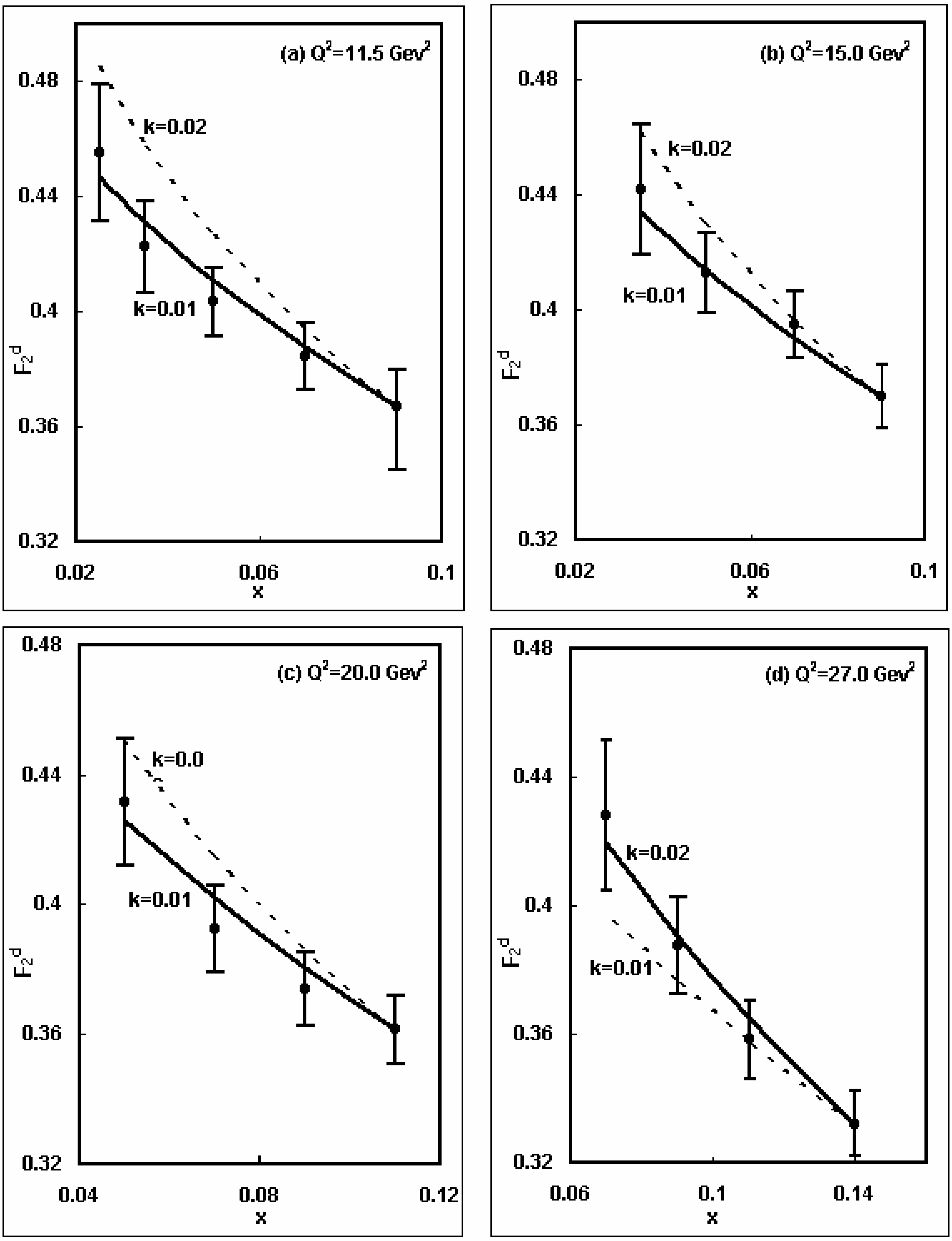

Figure 3